\documentclass[oldversion]{aa}  
\usepackage{graphicx}
\usepackage{txfonts}
\usepackage{natbib}
\begin{document}
\title{Spectral type dependent rotational braking and strong magnetic
  flux in three components of the late-M multiple system LHS~1070 }
\titlerunning{Spectral type dependent rotational braking}

   \author{A. Reiners
          \inst{1}\fnmsep\thanks{Emmy Noether Fellow}
          \and
          A. Seifahrt\inst{2,3}
          \and
          H.U. K\"aufl\inst{3}
          \and
          R. Siebenmorgen\inst{3}
          \and
          A. Smette\inst{4}
          }


   \institute{Universit\"at G\"ottingen, Institut f\"ur Astrophysik, Friedrich-Hund-Platz 1, D-37077 G\"ottingen, Germany\\
     \email{Ansgar.Reiners@phys.uni-goettingen.de}
     \and
     Universit\"at Jena, Astrophysikalisches Institut und Instituts-Sternwarte, Schillerg\"asschen 2, D-07745 Jena, Germany
     \and
     European Southern Observatory, Karl-Schwarzschild-Str. 2, D-85748 Garching, Germany
     \and
     European Southern Observatory, Alonso de C\'ordova 3107, Casilla 19001, Santiago 19, Chile
   }
   
   \date{Received 29 May 2007 / Accepted 14 June 2007}
   
 
  \abstract
  { We show individual high resolution spectra of components A, B, and
    C of the nearby late-M type multiple system LHS~1070. Component A
    is a mid-M star, B and C are known to have masses at the threshold
    to brown dwarfs. From our spectra we measure rotation velocities
    and the mean magnetic field for all three components individually.
    We find magnetic flux on the order of several kilo-Gauss in all
    components. The rotation velocities of the two late-M objects B
    and C are similar ($v\,\sin{i} = 16$\,km\,s$^{-1}$), the earlier A
    component is spinning only at about half that rate. This suggests
    weakening of net rotational braking at late-M spectral type, and
    that the lack of slowly rotating late-M and L dwarfs is real.
    Furthermore, we found that magnetic flux in the B component is
    about twice as strong as in component C at similar rotation rate.
    This indicates that rotational braking is not proportional to
    magnetic field strength in fully convective objects, and that a
    different field topology is the reason for the weak braking in low
    mass objects.}

  \keywords{Stars: low-mass, brown dwarfs -- Stars: magnetic fields --
    Stars: rotation -- Stars: individual: LHS 1070}

   \maketitle
%

\section{Introduction}

The rotation of Sun-like stars is braked following an empirically
determined braking law with $\mathbf{v} \propto t^{-0.5}$
\citep[][]{Skumanich72, Barnes07}. At several Gyr they have lost most
of their angular momentum and become slow rotators like the Sun.
Fully convective stars, however, are apparently not braked that much,
and field dwarfs of spectral type late-M or L rotate more rapidly than
their higher mass siblings \citep{Mohanty03}.

Stars are believed to lose angular momentum due to a magnetic stellar
wind after the first phase of star formation. Angular momentum loss
critically depends on the star's magnetic field and its geometry
\citep{Mestel84, Kawaler88, Chaboyer95, Sills00}. The mechanism of
magnetic field generation is of crucial importance since it determines
magnetic field strength, hence braking, and how it reacts on angular
velocity changes. Since the mechanism of magnetic field generation is
believed to change at the threshold to fully convective stars, it can
be expected that a change in the braking law appears as well.

The lack of slowly rotating objects of spectral type later than mid-M
could be a consequence of a change in the net braking. These objects,
however, are so faint that it is difficult to probe a sample that is
unbiased to luminosity effects, i.e. a luminosity limited sample will
always contain more bright (hence young and less braked) low mass
objects than fainter ones.

The close hierarchical system LHS~1070 (GJ~2005) is an ideal probe for
rotational evolution of late-M objects. Its individual components can
be spatially resolved; the A component is a mid-M star (M5.5) while
the two fainter components B and C are cooler (around M9).  Measuring
the astrometric orbit of B and C, \citet{Leinert00} determined masses
at the limit to brown dwarfs.

In this paper, we measure rotation velocities and magnetic flux of
LHS~1070 A, B, and C individually. Because all components are probably
coeval, this gives direct insight in spectral type dependent braking and
magnetic field generation.

\section{Data}

\begin{figure*}
  \centering
  \resizebox{\hsize}{!}{\includegraphics[clip=,bbllx=0,bblly=247,bburx=648,bbury=468]{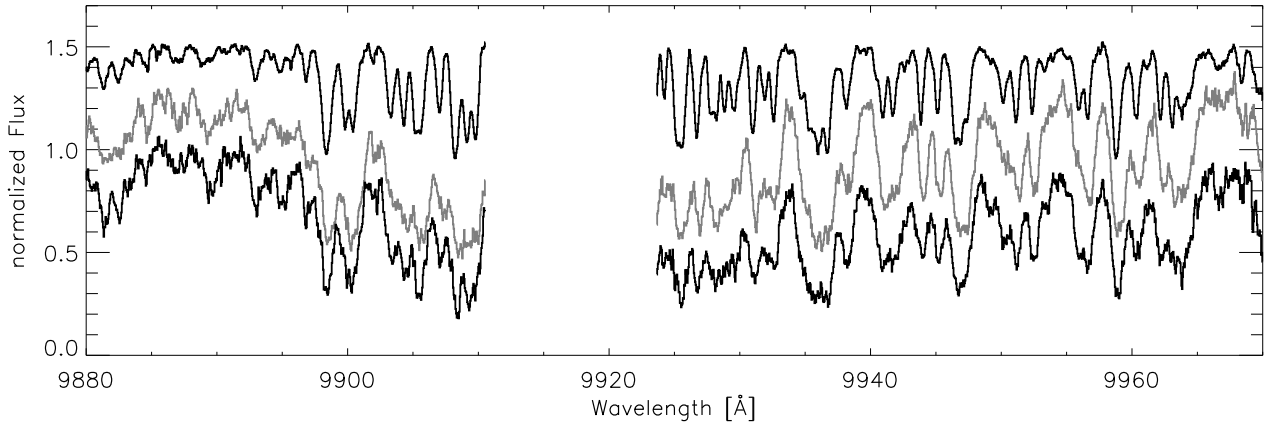}}
  \caption{\label{fig:Data}A part of our CRIRES data in the three
    components LHS 1070 A, B, and C (from top to bottom), smoothed by
    a 3 pixel box-car.}
\end{figure*}

In 2006, the three components LHS~1070\,A, B, and C were nearly
exactly lined up on the sky, a most suitable configuration for
simultaneous long-slit spectroscopy. We observed the triple in the
night of October 09, 2006 during a science verification run (Prog.ID
60.A-9078) of the newly commissioned AO-fed high resolution NIR
spectrograph CRIRES \citep{CRIRES} at Antu (UT1) of the VLT on Cerro
Paranal.

Separations of LHS~1070\,AB and AC are about 1.35\arcsec and
1.75\arcsec, respectively. The spectra are spatially well resolved
with only minimal overlap of the B and C components. Observations were
carried out in order 57/0/i with on-target AO, total exposure time of
was 480s. We chose a slitwidth of 0.4\arcsec, i.e. a nominal resolving
power of 50\,000.  Data reduction followed the standard procedure of
dark subtraction and flatfielding.  After proper background
subtraction to remove the OH airglow emission the spectra where
extracted using an adaption of the optimal extraction algorithm of
\citet{robertson}.  The chosen spectral window is clear of telluric
lines on a 2\% level.  Thus we do not need to use a standard star for
correction. Instead we determined the instrument response from the
flatfield spectra and removed the remaining tilt by minimal
rectification of the pseudo-continuum.

Wavelength calibration was performed using a ThAr spectrum in
comparison to an updated line list (Kerber, priv.comm.). Flux
contamination of the B and C components was found to be much smaller
than the intrinsic S/N ratio.  Thus we do not see any reason to doubt
on the spectral purity of any of the components in LHS~1070 for the
subsequent analysis. A part of our data is shown in
Fig.\,\ref{fig:Data}, at 9920\,\AA\ the spectrum falls on a gap
between two of the four chips.

\section{Analysis and Results} 

\subsection{Method}

Our observations were carried out in the Wing-Ford band of molecular
FeH (starting at 9900\,\AA), which shows sharp absorption features in
mid- to late-M dwarfs.  Some of the lines are strongly magnetically
sensitive while some are not \citep{Reiners06b}. This makes the region
ideal for the analysis of rotational broadening and the measurement of
magnetic fields in ultra-cool objects. Our analysis follows the
strategy used by \citet{Reiners07a} and \citet{Reiners07b}.

We use spectra of two template M-stars, one without any signs of
Zeeman broadening, and the other showing strong effects of Zeeman
broadening. First, the template spectra are scaled according to an
optical depth scaling law to fit the target's absorption strength.
Then, the two template spectra are artificially broadened to match the
targets rotation velocity.  Finally, we search for the linear
combination of the two template spectra that simultaneously fits the
magnetically sensitive and the magnetically insensitive lines. From
the interpolation parameter, we determine $Bf$, the product of the
mean magnetic field and the filling factor, assuming a linear relation
between Zeeman broadening and $Bf$ \citep[see][]{Reiners06b,
  Reiners07a}.

As template spectra, we use data of GJ~1002 (M5.5) and Gl~873 (M3.5)
taken with HIRES at Keck observatory\footnote{We thank G.  Basri for
  providing the template spectra.}, both stars have $v\,\sin{i} \le
3$\,km\,s$^{-1}$. Because the template spectra were taken at lower
spectral resolving power ($R = 31\,000$), we applied a Gaussian
broadening function to our data in order to match the resolution of
the templates. This is a crucial step since it sets the zero-point for
the calibration of rotational broadening. In order to check our
results, we carried out the analysis with a different set of templates
observed at higher resolution. We used a spectrum of GJ~1227 (M4.5,
$v\,\sin{i} \le 3$\,km\,s$^{-1}$) for the field-free template, and a
sunspot spectrum for the spectrum in the presence of a magnetic
field\footnote{\texttt{ftp://ftp.noao.edu}}. The spectrum of GJ~1227
was also taken with HIRES at Keck but at higher resolving power ($R
\sim 80\,000$). The template spectra were degenerated in order to
match the resolution of our CRIRES data. We find that both strategies
provide highly consistent results on the values of $v\,\sin{i}$ for
all three components of LHS~1070, i.e. differences are within
1\,km\,s$^{-1}$. For the following, we rely on the low resolution set
of templates because they provide a much more consistent data set, and
their resolution is entirely sufficient for our purposes.

\subsection{Rotation}

\begin{figure}
  \centering
  \resizebox{.9\hsize}{!}{\includegraphics[clip=,bbllx=0,bblly=257,bburx=648,bbury=446]{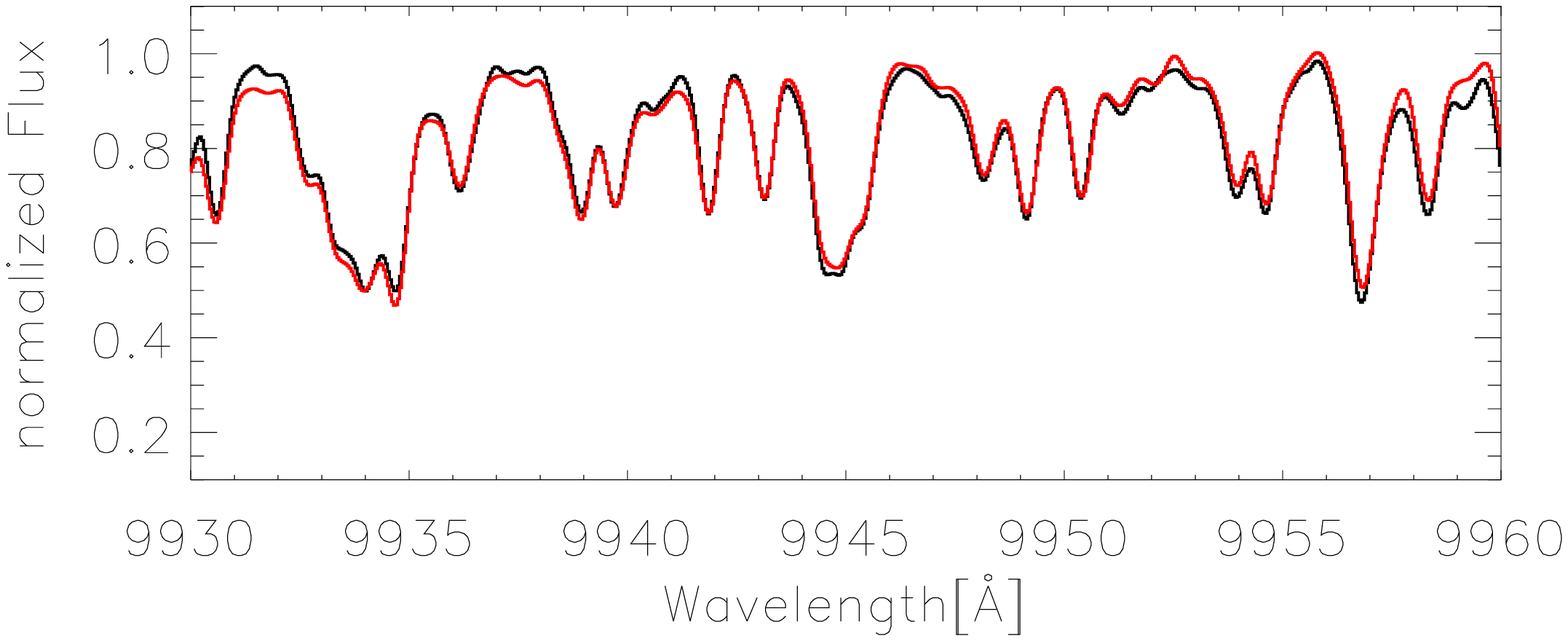}}
  \resizebox{.9\hsize}{!}{\includegraphics[clip=,bbllx=0,bblly=257,bburx=648,bbury=446]{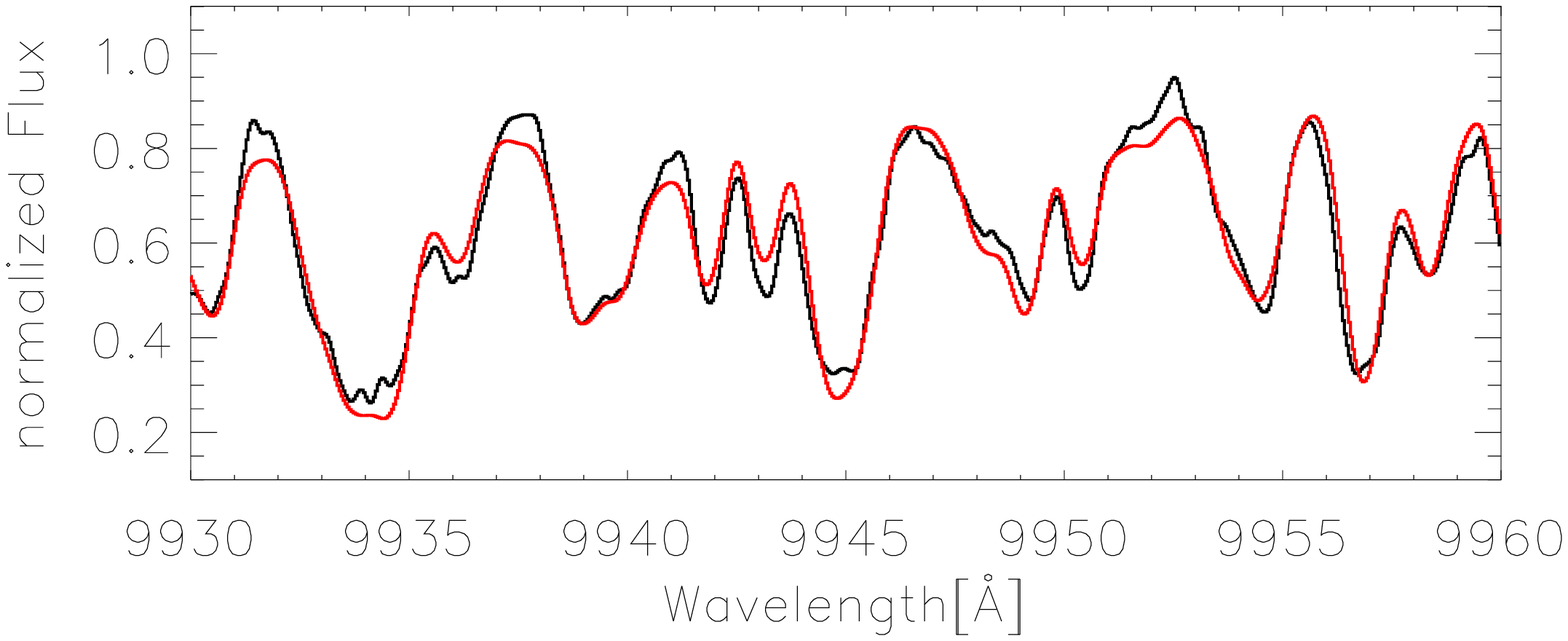}}
  \resizebox{.9\hsize}{!}{\includegraphics[clip=,bbllx=0,bblly=190,bburx=648,bbury=446]{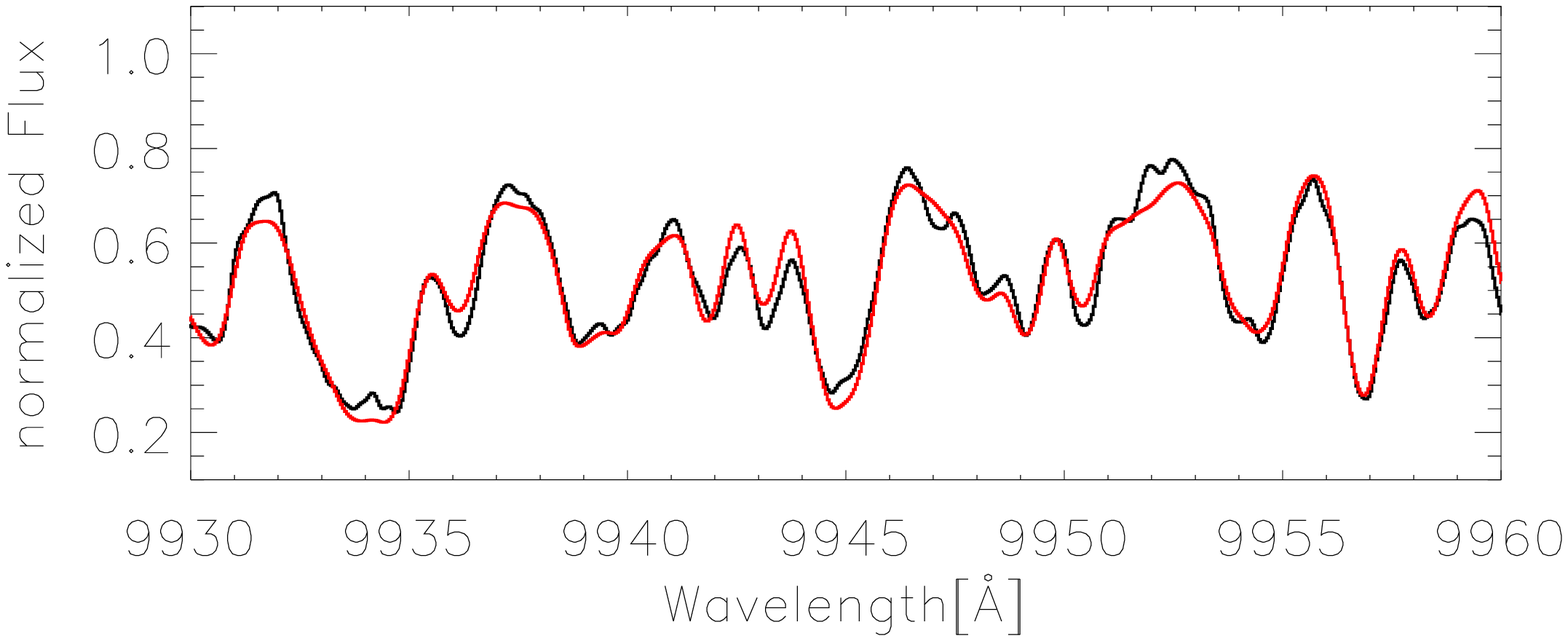}}
  \caption{\label{fig:vsinifit}Data (black) and fit (red) in the
    entire region we used for the fit in the three components
    LHS~1070, A, B, and C (from top to bottom). The data was
    degenerated in resolution to match the resolution of the template
    spectra ($R = 31\,000$). }
\end{figure}

We show our best fits over the 30\,\AA-wide region used for fitting as
red lines in all three components of LHS~1070 in
Fig.\,\ref{fig:vsinifit}. Smoothing our spectra to match the lower
resolution of the template spectra strongly reduced the effective
noise level. S/N ratios per resolution element are now on the order of
70 for LHS~1070 A, 35 and 30 for LHS~1070 B and C, respectively.
Differences in projected rotation velocity can be distinguished by eye
for example at 9939\,\AA\ where two neighbored lines are just
separated in the A-component but already blended in components B and
C. The two latter objects do not show any obvious difference in the
width of individual lines.

We derive projected rotation velocities of $v\,\sin{i} = 8$, 16, and
16\,km\,s$^{-1}$ in LHS~1070 A, B, and C, respectively. The overall
fit quality is rather good particularly in the A-component of the
system. In the two late-M components B and C, the fit has some
weaknesses which we attribute to the much stronger absorption that may
not be perfectly captured by the optical depth scaling. However, for
our analysis of $v\,\sin{i}$ the quality is entirely sufficient. A
formal $\chi^2$ analysis shows that the uncertainty in $v\,\sin{i}$ is
on the order of 1\,km\,s$^{-1}$ for the three objects. The absolute
uncertainty is somewhat higher since the zero-point depends on the
resolution we adopt for the template spectra and our target spectra.
We estimate that absolute uncertainties are on the order of
3\,km\,s$^{-1}$, but that does not affect the comparison of rotation
velocities between the three components of LHS~1070. \cite{Basri95}
have measured $v\,\sin{i}$ in a spatially unresolved spectrum of
LHS~1070. They find $v\,\sin{i} = 8\pm2$\,km\,s$^{-1}$ in excellent
agreement with our result for LHS~1070\,A, which dominates the
spatially unresolved spectrum.

\subsection{Magnetic flux}

Magnetic flux was determined by searching for a linear interpolation
between a field free template spectrum and a template spectrum
affected by magnetic Zeeman broadening. For the latter, we use a
spectrum of Gl~873 and we adopt a magnetic flux of $Bf = 3.9$\,kG
\citep[see][]{Reiners07a}. This implies that our results cannot exceed
values of $\sim 4$\,kG.

\begin{figure}
  \centering
  \resizebox{.9\hsize}{!}{\includegraphics[clip=,bbllx=0,bblly=257,bburx=648,bbury=446]{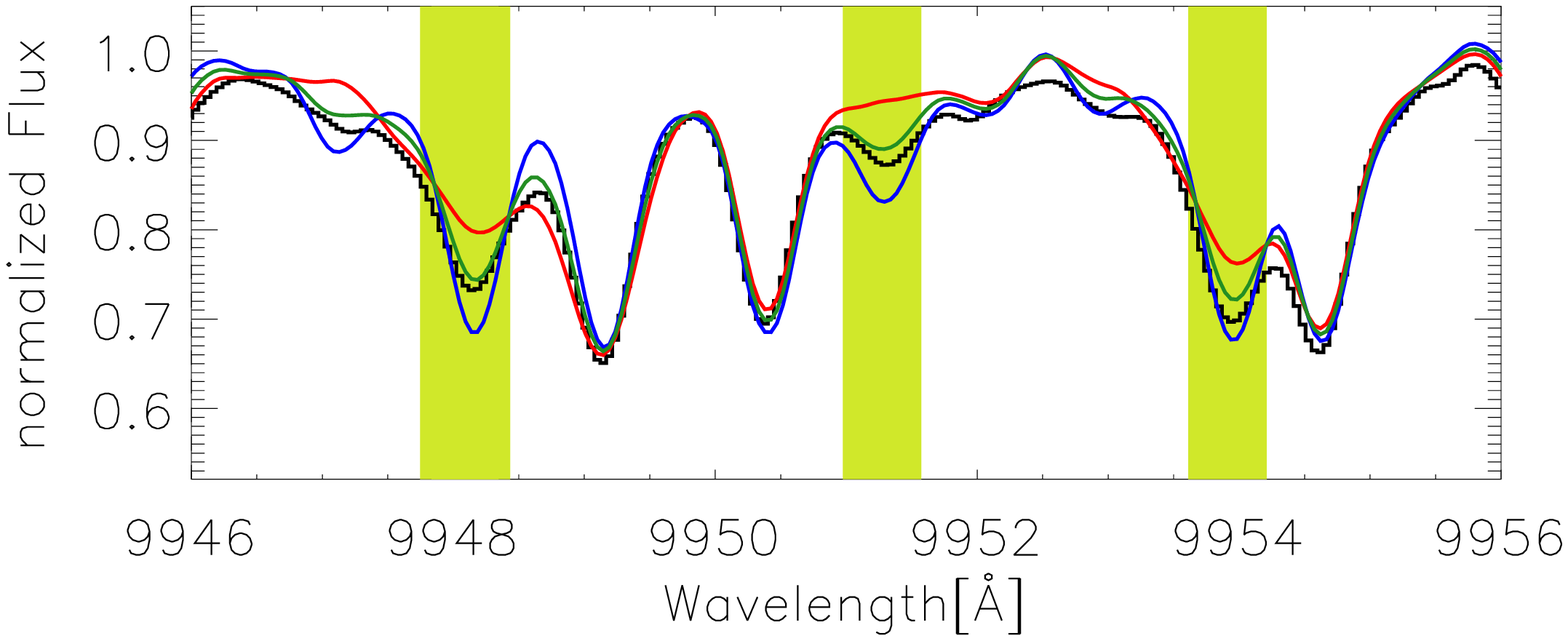}}
  \resizebox{.9\hsize}{!}{\includegraphics[clip=,bbllx=0,bblly=257,bburx=648,bbury=446]{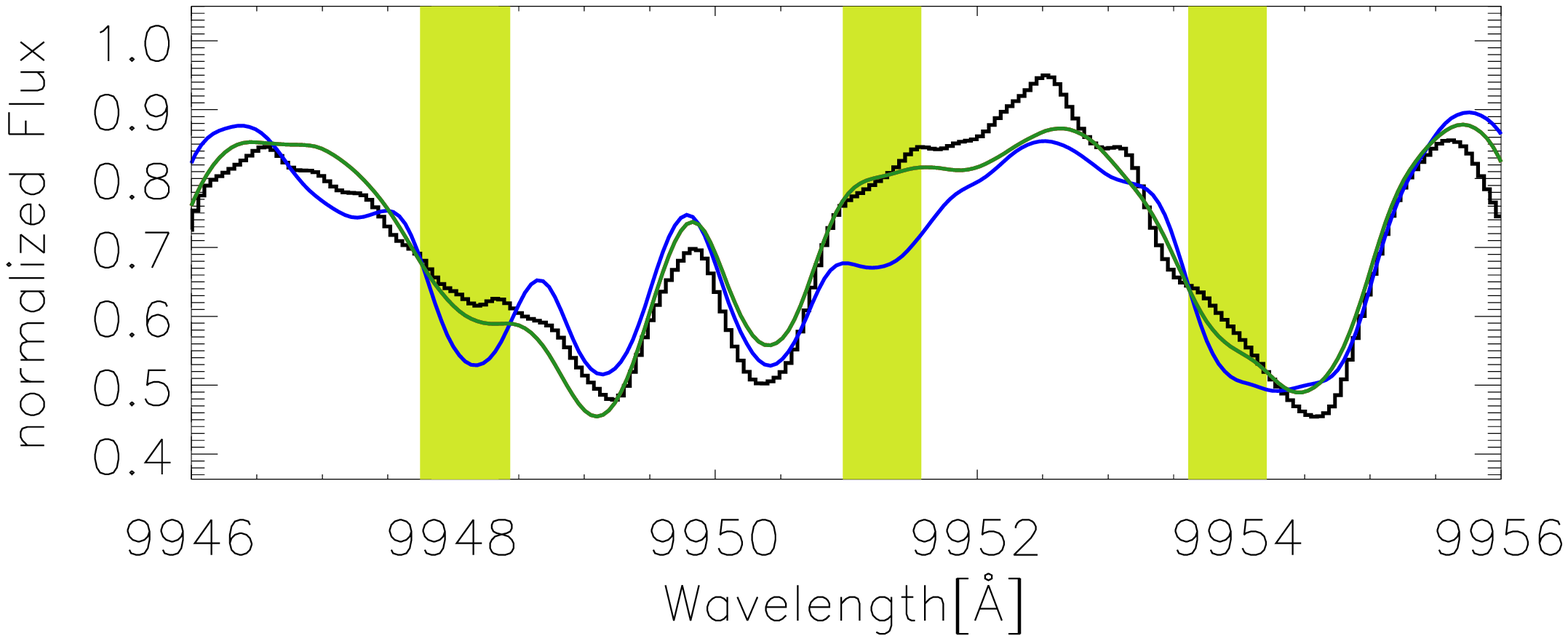}}
  \resizebox{.9\hsize}{!}{\includegraphics[clip=,bbllx=0,bblly=190,bburx=648,bbury=446]{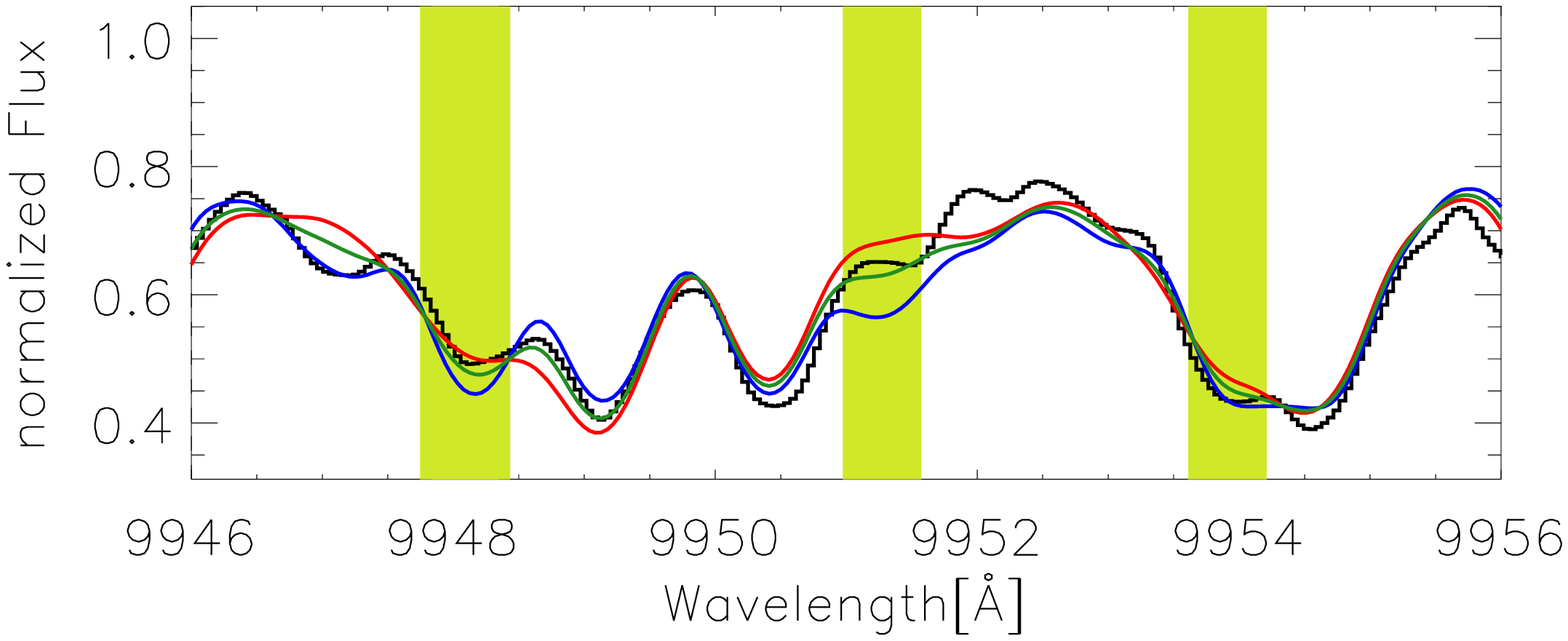}}
  \caption{\label{fig:bffit}Portion of the spectral range that shows
    the effect of magnetic flux. Green bars mark magnetically most
    sensitive lines. Top to bottom: LHS~1070 A, B, and C. Black lines
    are the target spectra, red lines mark spectra affected by strong
    magnetic flux (Gl\,873), blue lines spectra without magnetic flux
    (GJ~1002). Green lines are our best fits. Note that in the B
    component the red line is covered by the green one. }
\end{figure}

At every wavelength, the interpolation lies between the two template
spectra. In magnetically insensitive regions, the difference is very
small while at wavelengths that are affected by Zeeman broadening, a
difference proportional to the magnetic flux occurs. We show a
representative region of our spectra at the wavelength range
9946--9956\,\AA\ in Fig.\,\ref{fig:bffit}. Here, we highlight three
absorption lines that are particularly sensitive to Zeeman broadening.
We plot the target spectra in black and overplot the (rotationally
broadened) template spectra of GJ\,1002 (no magnetic flux) in blue,
and the one of Gl~873 (strong flux) in red. The interpolation that
best fits the data is overplotted in green.

A very accurate fit is achieved in the spectrum of LHS~1070\,A, from
which we derive a magnetic flux of 2.0\,kG. The fit quality in the
spectrum of LHS~1070\,B is not as accurate. However, it appears that
the two sensitive absorption lines at 9948\,\AA\ and 9954\,\AA\ are
strongly smeared out, possibly even more than in the template spectrum
of Gl~873. This indicates very strong magnetic flux, and we adopt a
value of $Bf \sim 4$\,kG for component B (note that the red line is
covered by the green fit in this panel). The presence of strong
magnetic flux can also be seen by comparing the spectrum of
LHS~1070\,B to the one of LHS~1070\,C in the third panel. Here, the
fit quality is much higher again, we deduce a magnetic flux of 2\,kG
as in the A-component. Comparing this spectrum to the one of
LHS~1070\,B (particularly the lines at 9948\,\AA and 9954\,\AA), it
appears that the magnetically sensitive absorption lines are much
weaker in LHS~1070\,B than they are in LHS~1070\,C.

As discussed in \citet{Reiners06b, Reiners07a}, our lack of knowledge
about the Zeeman splitting patterns of the FeH molecule causes the
main uncertainty in the determination of magnetic flux using this
molecule as a tracer. On an absolute scale, we estimate our
uncertainties in magnetic flux on the order of a kilo-Gauss. However,
the main result of our magnetic flux analysis is already seen in a
rough comparison of the three spectra: All three components show a
substantial effect of magnetic flux on the spectral lines, and the
magnetic flux in LHS~1070\,B is stronger than the flux in LHS~1070\,A
and LHS~1070\,C.

\section{Summary and Conclusions}

\begin{table}
  \caption{\label{tab:results}Parameter of the three components 
    LHS~1070\,A, B, and C. Spectral types are from \citet{Leinert00}.}
  \begin{tabular}{lccccc}
    \hline
    \hline
    \noalign{\smallskip}
    Component & SpType & $V$ & $J$ & $v\sin{i}$ & $Bf$\\
    & & & & [km/s$^{-1}$] & [kG]\\
    \noalign{\smallskip}
    \hline
    \noalign{\smallskip}
    LHS 1070 A & M 5.5 & 15.42 & 9.25 & $ 8$ & 2.0 \\
    LHS 1070 B & M 8.5 &       &      & $16$ & 4.0 \\
    LHS 1070 C & M 9.0 &       &      & $16$ & 2.0 \\
    \noalign{\smallskip}
    \hline
    \hline
  \end{tabular}
\end{table}

We have isolated high-resolution spectra of the three components of
the multiple M-object system LHS~1070. From each spectrum, we have
determined the projected rotation velocity $v\,\sin{}$ and the mean
magnetic field over the entire star. The results of our analysis are
summarized in Table\,\ref{tab:results}. We find a projected rotational
velocity of $v\,\sin{i} = 8$\,km\,s$^{-1}$ in the hottest component,
LHS~1070\,A, while rotation in LHS~1070\,B and C is about twice as
rapid. Under the reasonable assumption that the entire system LHS~1070
was formed at the same time, we are facing three different objects
that have evolved from the same initial conditions. Disc orientations
in pre-main sequence stars \citep{Monin06} and orbit orientations in
multiple systems \citep{Sterzik02} are partially correlated, and
\cite{Bate00} find that strong misalignments are unlikely in binaries
with separations $\le 100$\,AU.  Thus, we may assume that the
inclination angles $i$ of LHS~1070\,A, B and C are comparable
\citep[see also][]{Leinert01}.  As long as they suffer the same
rotational braking, we would expect comparable rotation velocities in
all components. We observe comparable rotation rates in the components
B and C, which are of very similar spectral type.  However, the
rotation velocity in LHS~1070\,A is about a factor of two smaller.
This can probably be attributed to a difference in the rotational
braking during their evolution.

\cite{Mohanty03} showed that in the field the fraction of slow
rotators ($v\,\sin{i} < 3$\,km\,s$^{-1}$) later than spectral type M6
is much smaller than among early-M dwarfs, which generally rotate very
slowly \citep{Delfosse98}. However, the sample of \cite{Mohanty03}
could be biased to bright (hence young) objects that are still more
rapidly rotating than their older (hence fainter) siblings.
\citet{Reiners06a} found slow rotation in a mid/late-M subdwarf, and
very rapid rotation ($v\,\sin{i} = 65$\,km\,s$^{-1}$) in a late-L
subdwarf. Since late-type subdwarfs are probably the oldest relics of
the early galaxy having spun down for several Gyr, this is another
argument for less efficient rotational braking at later spectral type.

\begin{figure}
  \centering
  \resizebox{.9\hsize}{!}{\includegraphics{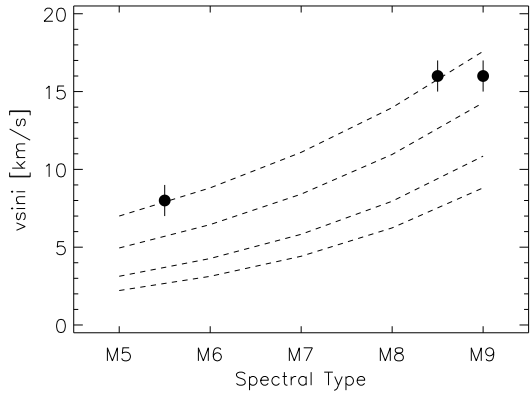}}
  \caption{\label{fig:vsinispec}Projected rotation velocities of
    LHS~1070\,A, B, and C. We overplot rotation velocities according
    to a modified Skumanich braking law at 1, 2, 5, and and 10\,Gyr
    (top to bottom).}
\end{figure}

In LHS~1070, we see three objects of different spectral type that are
probably of same age. We plot the projected rotation velocities of
LHS~1070\,A, B, and C in Fig.\,\ref{fig:vsinispec}. As stated above,
the two later objects show higher rotation rates than the earlier
A-component. One explanation may be a difference in the initial
conditions, but together with \cite{Mohanty03} this would suggest that
later objects in general start off rotating more rapidly, which is
unlikely. On the other hand, the rotation activity connection seems to
vanish around late-M spectral type. This is likely connected to a
modified wind-law and consequently to different rotational braking.

Detailed simulations of rotational braking were performed by
\citet{Chaboyer95, Sills00}. They give a prescription for a
wind-braking law in stars at given angular velocity, and they show
that the braking law in fully convective stars requires different
treatment \citep[to fit observations in the Hyades they had to tune
their value of $\omega_{\rm{crit}}$;][]{Sills00}. In this work, we do
not intend to give a full description of the braking law in fully
convective objects. Instead, we assume that the net braking law in
LHS~1070\,A is still Skumanich-type \citep[$v \propto
t^{-0.5}$,][]{Skumanich72}, and that the exponent of the braking law
is gradually decreasing towards lower temperatures.

For this purpose, we calculate the rotational evolution starting at an
age of 10\,Myr and an initial rotation velocity of 70\,km\,s$^{-1}$.
For our spectral type dependent braking law, we chose $v \propto
t^{-\alpha}$ with gradually decreasing $\alpha = 0.5$--0.3 in the
spectral range M5--M9. Our results are overplotted as dashed lines in
Fig.\,\ref{fig:vsinispec}. Four lines for ages of 1, 2, 5, and 10\,Gyr
are shown (top to bottom). The assumed spectral type dependent scaling
law can reproduce the difference in rotation velocities as measured in
LHS~1070, and at the same time generate the rise in minimum rotational
velocities consistent to measurements in late-M stars
\citep{Mohanty03}. We thus conclude that braking in late-M stars is
spectral-type dependent, and that the lack of slowly rotating late-M
dwarfs is not a selection effect due to a bias towards young rapid
rotators.

The age estimated from our approach is on the order of 1\,Gyr.
\citet{Basri95} found a high velocity component perpendicular to the
galactic plane that suggests higher age. We think that this point
deserves further attention. Although age estimates from space velocity
are rather uncertain, this may hint to an absolute scale of the
braking law different to the one used here.

All three components of LHS~1070 show strong magnetic flux suggesting
that in fully convective objects, the field strength does not strongly
depend on rotation. \citet{Leinert00} found signs of magnetic activity
in components A and B, but not in C. This is consistent with our
results in the sense that the magnetic flux in LHS~1070\,B is much
stronger than the one in LHS~1070\,C.  Nevertheless, one expects that
the C component also exhibits H$\alpha$ emission, which may still be
below the detection limit of \citet{Leinert00}.

Interestingly, the difference in magnetic flux between LHS~1070\,B and
C does apparently not influence rotational braking. It is generally
believed that angular momentum loss depends on magnetic field strength
with an exponent given by the field geometry \citep{Mestel84}.  This
coupling, however, depends on the field generation mechanism
\citep{Kawaler88}, and angular momentum loss becomes saturated at a
certain level \citep{Chaboyer95}. According to \citet{Chabrier06},
large scale magnetic fields of equipartition strength (a few kG) can
be generated by an $\alpha^2$ dynamo. These authors predict the field
topology to differ from an organized dipole field, which could explain
our result that rotational braking does not grow with magnetic field
strength. It would imply that field geometry is the reason for lower
rotational braking in low mass objects.
 
\begin{acknowledgements}
  We like to thank the CRIRES science verification team, for their
  work on CRIRES and for the execution of the observations, and the
  referee, Subu Mohanty, for a very constructive report. AR
  acknowledges financial support through an Emmy Noether Fellowship
  from the Deutsche Forschungsgemeinschaft under DFG RE 1664/4-1.
\end{acknowledgements}

\end{document}